\RequirePackage{fix-cm}
\documentclass[twocolumn,epjc3]{svjour3}  
\RequirePackage{graphicx}
 \RequirePackage{mathptmx}      
\RequirePackage[numbers,sort&compress]{natbib}
\RequirePackage[displaymath, mathlines]{lineno} 
\RequirePackage{amssymb} 
\RequirePackage{amsmath}

\journalname{Eur. Phys. J. C}
\begin{document}

\title{Searches for Relativistic Magnetic Mono\-poles in IceCube
}


\onecolumn 
\author{IceCube Collaboration: M.~G.~Aartsen\thanksref{Adelaide}
	\and K.~Abraham\thanksref{Munich}
	\and M.~Ackermann\thanksref{Zeuthen}
	\and J.~Adams\thanksref{Christchurch}
	\and J.~A.~Aguilar\thanksref{BrusselsLibre}
	\and M.~Ahlers\thanksref{MadisonPAC}
	\and M.~Ahrens\thanksref{StockholmOKC}
	\and D.~Altmann\thanksref{Erlangen}
	\and T.~Anderson\thanksref{PennPhys}
	\and I.~Ansseau\thanksref{BrusselsLibre}
% ...
}

\authorrunning{IceCube Collaboration}
\thankstext{a}{Corresponding author: anna.pollmann@uni-wuppertal.de} 
\thankstext{b}{Corresponding author: jposselt@icecube.wisc.edu}
\thankstext{c}{Earthquake Research Institute, University of Tokyo, Bunkyo, Tokyo 113-0032, Japan}
\thankstext{d}{NASA Goddard Space Flight Center, Greenbelt, MD 20771, USA}
\institute{
% ...
Department of Physics, University of Adelaide, Adelaide, 5005, Australia \label{Adelaide}
% ...
\and Universit\'e Libre de Bruxelles, Science Faculty CP230, B-1050 Brussels, Belgium \label{BrusselsLibre}
% ...
\and Dept.~of Physics and Astronomy, University of Canterbury, Private Bag 4800, Christchurch, New Zealand \label{Christchurch}
% ...
\and Erlangen Centre for Astroparticle Physics, Friedrich-Alexander-Universit\"at Erlangen-N\"urnberg, D-91058 Erlangen, Germany \label{Erlangen}
% ...
\and Dept.~of Physics and Wisconsin IceCube Particle Astrophysics Center, University of Wisconsin, Madison, WI 53706, USA \label{MadisonPAC}
% ...
\and Technische Universit\"at M\"unchen, D-85748 Garching, Germany \label{Munich}
% ...
\and Oskar Klein Centre and Dept.~of Physics, Stockholm University, SE-10691 Stockholm, Sweden \label{StockholmOKC}
% ...
\and Dept.~of Physics, Pennsylvania State University, University Park, PA 16802, USA \label{PennPhys}
% ...
\and DESY, D-15735 Zeuthen, Germany \label{Zeuthen}
} 

\date{Received: date / Accepted: date}

\maketitle

\twocolumn

\setlength\linenumbersep{5pt}
\renewcommand\linenumberfont{\normalfont\tiny\sffamily}
%


\begin{abstract} 

Various extensions of the Standard Model motivate the existence of stable magnetic monopoles that could have been created during an early high-energy epoch of the Universe. These primordial magnetic monopoles would be gradually accelerated by cosmic magnetic fields and could reach high velocities that make them visible in Cherenkov detectors such as IceCube.

Equivalently to electrically charged particles, magnetic mono\-poles produce direct and indirect Cherenkov light while traversing through matter at relativistic velocities. 

This paper describes  searches for relativistic ($v\geq0.76\;c$) and mildly relativistic ($v\geq0.51\;c$) mono\-poles,  each using one year of data taken in 2008/09 and 2011/12 respectively. 
No mono\-pole candidate was detected. 
For a velocity above $0.51 \; c$ the mono\-pole flux is constrained down to a level of $1.55 \cdot 10^{-18} \; \textrm{cm}^{-2} \textrm{s}^{-1} \textrm{sr}^{-1}$. 
This is an improvement of almost two orders of magnitude over previous limits. 

\end{abstract}


\section{Introduction}\label{sec:Intro}

In Grand Unified Theories (GUTs) the existence of magnetic mono\-poles follows from general principles~\cite{THooft74,Polyakov74}. 
Such a theory is defined by a non-abelian gauge group that is spontaneously broken at a high energy to the the Standard Model of particle physics \cite{Guth80}. 
The condition that the broken symmetry contains the electromagnetic gauge group $\mathrm{U(1)_{\rm EM}}$ is sufficient for the existence of magnetic mono\-poles \cite{Polchinski04}.
Under these conditions the monopole is predicted to carry a magnetic charge $g$ governed by Dirac's quantization condition \cite{Dirac1931} 
 \begin{linenomath}
 	\begin{equation}
 		\label{eq:Dirac}
 		g=n \cdot g_D = n \cdot \frac{e}{2\alpha}
 	\end{equation}
 \end{linenomath}
 where $n$ is an integer, $g_D$ is the elemental magnetic charge or Dirac charge, $\alpha$ is the fine structure constant, and $e$ is the elemental electric charge. 

In a given GUT model the monopole mass can be estimated by the unification scale $\Lambda _{\textrm{GUT}}$ and the corresponding value of the running coupling constant $\alpha _{\textrm{GUT}}$ as $M c^2 \gtrsim {\Lambda _{\textrm{GUT}}}/{\alpha_{\textrm{GUT}}}$. Depending on details of the GUT model, the monopole mass can range from $10^7 \, \textrm{GeV}/c^2$ to $10^{17}\, \textrm{GeV}/c^2$ \cite{Preskill84, Wick03}. In any case, GUT mono\-poles are too heavy to be produced in any existing or foreseeable accelerator. 

After production in the very early hot universe, their relic abundance is expected to have been exponentially diluted during inflation. 
However, mono\-poles associated with the breaking of intermediate scale gauge symmetries may have been produced in the late stages of inflation and reheating in some models~\cite{Dar2006, Sakellariadou2008}. There is thus no robust theoretical prediction of monopole parameters such as mass and flux, nevertheless an experimental detection of a mono\-pole today would be of fundamental significance.

In this paper we present results for mono\-pole searches with the IceCube Neutrino telescope covering a large velocity range. Due to the different light-emitting mechanisms at play, we present two analyses, each optimized according to their velocity range: 
highly relativistic mono\-poles with $v\geq0.76\,c$ and mildly relativistic mono\-poles with $v\geq0.4\,c$. 
The highly relativistic mono\-pole analysis was performed with IceCube in its 40-string configuration while the mildly relativistic mono\-pole analysis uses the complete 86-string detector.

The paper is organized as follows. In section \ref{sec:IC} we introduce the neutrino detector IceCube and describe in section \ref{sec:Cher} the methods to detect magnetic mono\-poles with Cherenkov telescopes. We describe the simulation of magnetic mono\-poles in section \ref{sec:Sim}. The analyses for highly and mildly relativistic mono\-poles use different analysis schemes which are described in sections \ref{sec:IC40} and \ref{sec:IC86}. The result of both analyses and an outlook is finally shown in sections \ref{sec:result} to \ref{sub:outlook}.


\begin{figure}
	\centering
	\includegraphics[width=3in]{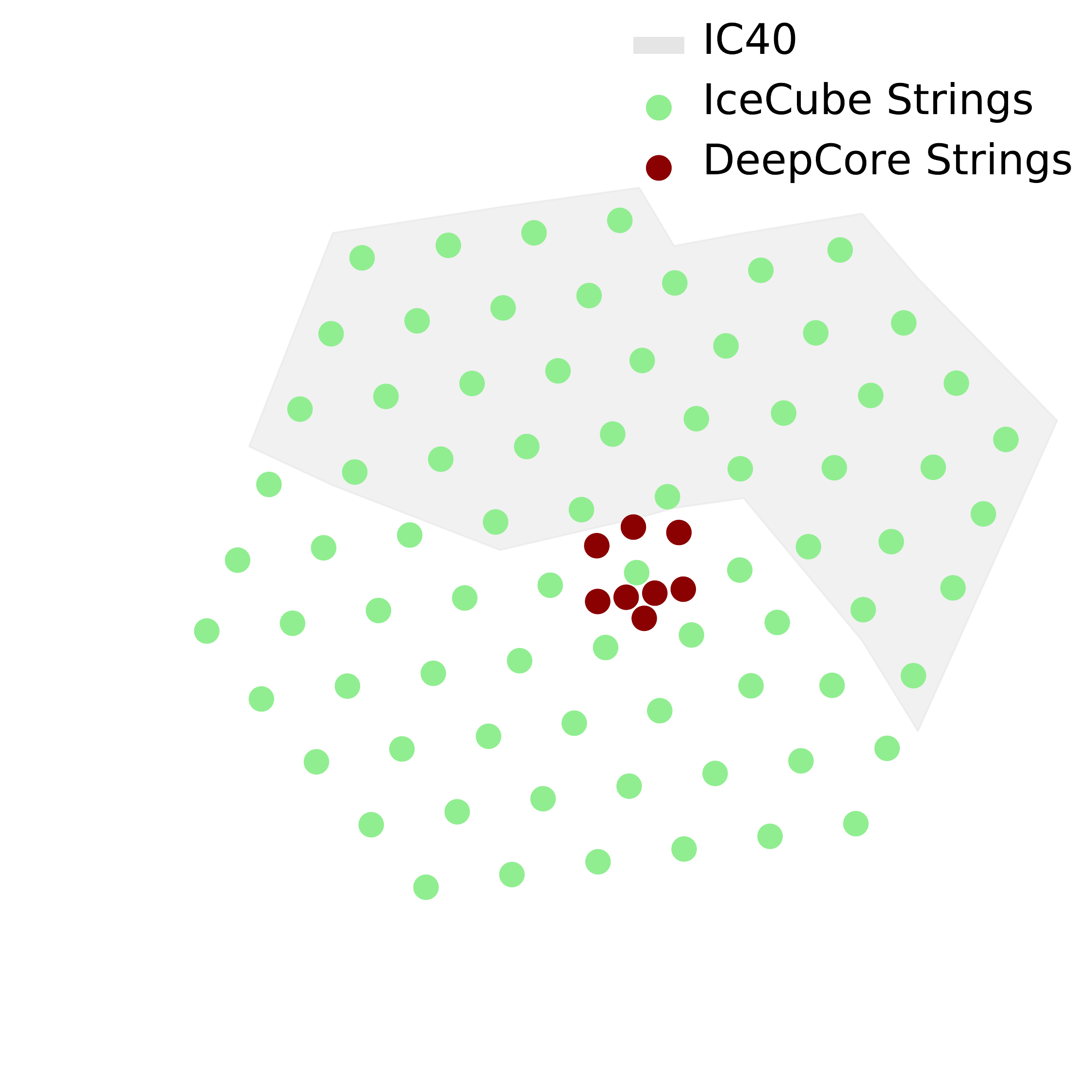} 
	\caption{A top view of the IceCube array. 
		The IC40 configuration consists of all strings in the upper gray shaded area. After completion in the end of 2010, IceCube consists of all 86 strings, called the IC86 configuration.
		DeepCore strings were excluded in the presented analyses.} 
	\label{fig:IC-configuration}
\end{figure}

\section{IceCube} \label{sec:IC}

The IceCube Neutrino Observatory is located at the geographic South Pole and consists of an in-ice array, IceCube \cite{IceCube06}, and a surface air shower array, IceTop \cite{IceTop13}, dedicated to neutrino and cosmic ray research, respectively. 
An aerial sketch of the detector layout is shown in Fig. \ref{fig:IC-configuration}.
	
IceCube consists of 86 strings with 60 digital optical modules (DOMs) each, deployed at depths between $1450\,\textrm{m}$ and $2450\,\textrm{m}$, instrumenting a total volume of one cubic kilometer. Each DOM contains a $25\;\textrm{cm}$ Hamamatsu photomultiplier tube (PMT) and electronics to read out and digitize the analog signal from the PMT \cite{IceCubePMT10}. 
The strings form a hexagonal grid with typical inter-string separation of $125\,\textrm{m}$ and vertical DOM separation of $17\,\textrm{m}$, except for six strings in the middle of the array that are more densely instrumented (with higher efficiency PMTs) and deployed closer together. 
These strings constitute the inner detector, DeepCore \cite{DeepCore12}. 
Construction of the IceCube detector started in December 2004 and was finished in December 2010, 
but the detector took data during construction. 
Specifically in this paper, we present results from two analyses, one performed with one year of data taken during 2008/09, when the detector consisted of 40 strings, called IC40, and another analysis with data taken during 2011/12 using the complete detector, called IC86. 

IceCube uses natural ice both as target and as radiator. 
The properties of light propagation in the ice must be measured thoroughly in order to accurately model the detector response. 
The analysis in the IC40 configuration of highly relativistic mono\-poles uses a six-parameter ice model \cite{AHA06} which describes the depth-dependent extrapolation of measurements of scattering and absorption valid for a wavelength of $400\,\textrm{nm}$. 
The IC86 analysis of mildly relativistic mono\-poles uses an improved ice model which is based on additional measurements and accounts for different wavelengths \cite{Spice13}. 

Each DOM transmitted digitized PMT waveforms to the surface.  The number of photons and their arrival times were then extracted from these waveforms.
The detector is triggered when a DOM and its next or next-to-nearest DOMs record a hit within a $1\, \mu \textrm{s}$ window. Then all hits in the detector within a window of $10\, \mu \textrm{s}$ will be read-out and combined into one event \cite{IceCubeDAQ09}. A series of data filters are run on-site in order to select potentially interesting events for further analysis, reducing at the same time the amount of data to be transferred via satellite. For both analyses presented here, a filter selecting events with a high number of photo-electrons ( $>\,650$ in the highly relativistic analysis and $>\,1000$ in the mildly relativistic analysis) were used. In addition filters selecting up-going track like events are used in the mildly relativistic analysis.

After the events have been sent to the IceCube's computer farm, they undergo some standard processing, such as the removal of hits which are likely caused by noise and basic reconstruction of single particle tracks via the LineFit algorithm \cite{ImprLineFit14}.
This reconstruction is based on a 4-dimensional (position plus time) least-square fit which yields an estimated direction and velocity for an event.

The analyses are performed in a blind way 
 by optimizing the cuts to select a possible monopole signal on simulation and one tenth of the data sample (the burn sample). The remaining data is kept untouched until the analysis procedure is fixed \cite{Roodman03}.
In the highly relativistic analysis the burn sample consists of all events re\-corded in August of 2008. In the mildly relativistic analysis the burn sample consists of every 10th 8-hour-run in 2011/12.


\begin{figure}
	
	\includegraphics[width=3in]{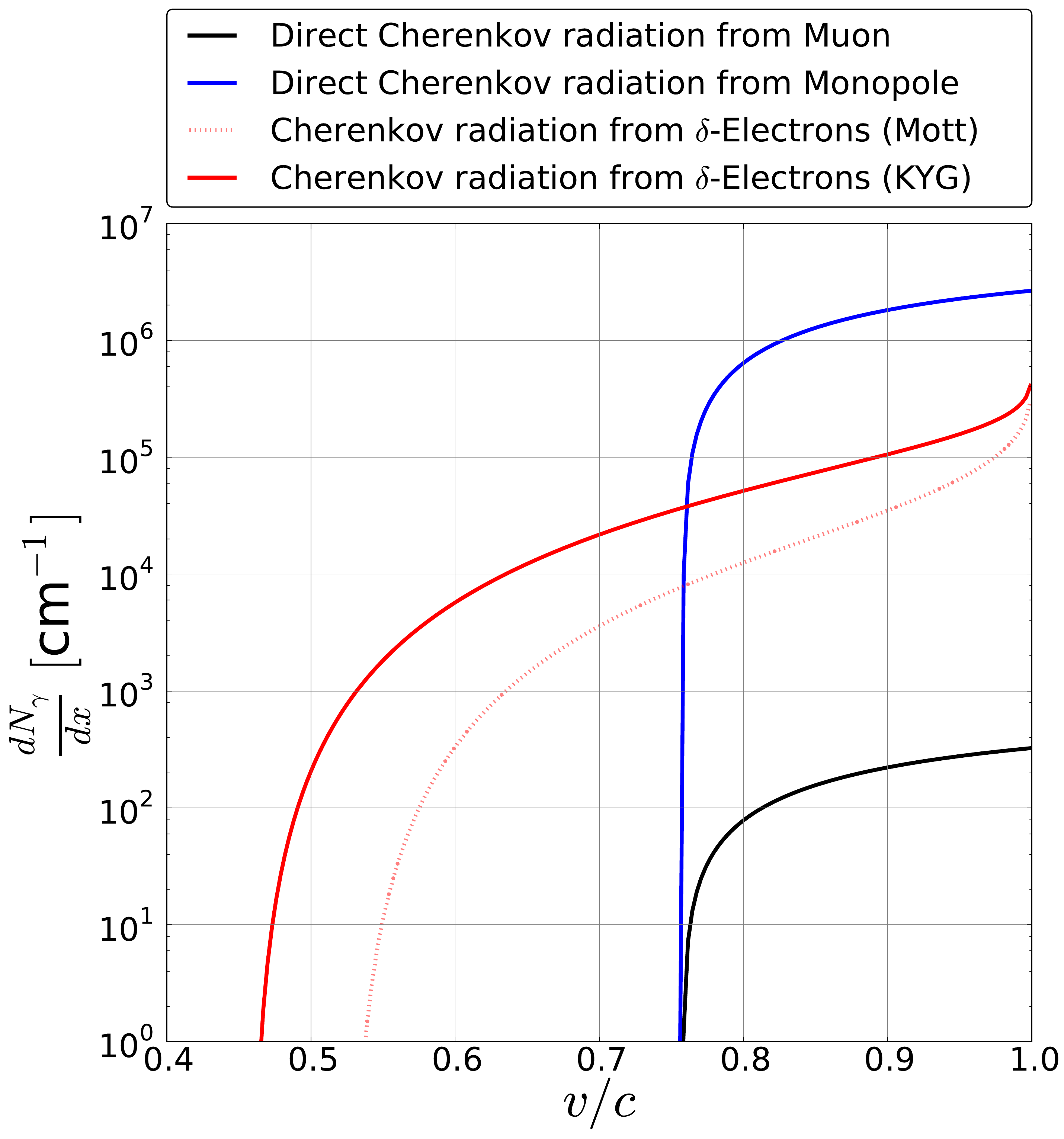}
	\caption{Number of photons per cm produced by a muon (black), a mono\-pole by direct Cherenkov light (blue), and mono\-poles by $\delta$-electrons. The photon yield per indirect Cherenkov light is shown using the KYG (red solid)  and, for comparison, the Mott (red dotted) cross section, used in one earlier monopole analysis \cite{Antares12}. 
		Light of wavelengths from $300\,\textrm{nm}$ to $600\,\textrm{nm}$ is considered here, covering the DOM acceptance of IceCube \cite{Spice13}
	} 
	\label{num_photons}
\end{figure}

\section{Monopole Signatures \label{sec:Cher}} 

Magnetic monopoles can gain kinetic energy through acceleration in magnetic fields.
This acceleration follows from a generalized Lorentz force law \cite{Moulin01} and is analogous to the
acceleration of electric charges in electric fields. The kinetic energy gained by a monopole of
charge \(g_D\) traversing a magnetic field \(B\) with coherence length \(L\) is $E \sim g_D BL\,$ 
\cite{Wick03}. This gives a gain of up to $10^{14}\,\textrm{GeV}$ of kinetic energy in intergalactic magnetic fields to reach relativistic velocities.
At such high kinetic energies magnetic monopoles can pass through the Earth while still having relativistic velocities when reaching the IceCube detector.

In the mono\-pole velocity range considered in these analyses, $v \geq 0.4\,c$ at the detector, 
three processes generate detectable light: direct Cherenkov emission by the  mono\-pole itself, indirect Cherenkov emission from ejected $\delta$-electrons and luminescence. 
Stochastical energy losses, such as pair production and photonuclear reactions, are neglected 
because they just occur at ultra-rel\-a\-tivis\-tic velocities.

An electric charge $e$ induces the production of Cherenkov light when its velocity $v$ exceeds the Cherenkov threshold $v_C=c/n_P\approx 0.76\,c$ where $n_P$ is the refraction index of ice. 
A magnetic charge $g$ moving with a velocity $\beta=v/c$ produces an electrical field whose strength is  proportional to the particle's velocity and charge. At velocities above $v_C$, Cherenkov light is produced analogous to the production by electrical charges \cite{Tamm37}
in an angle $\theta$ of 
 \begin{linenomath}
 	\begin{equation}
 	\label{EQ:CherCone}
 	\cos \theta = \frac{1}{n_P\,\beta}
 	\end{equation}
 \end{linenomath}
The number of Cherenkov photons per unit path length $x$ and wavelength $\lambda$ emitted by a monopole with 
one 
magnetic charge $g=g_D$ 
can be described by the usual Frank-Tamm formula \cite{Tamm37} for a particle with effective charge $Ze \rightarrow 
g_D 
n_P$ \cite{Tompkins}
\begin{linenomath}
\begin{equation}
\label{eq:DN}
\frac{d^2 N_{\gamma}}{dx d\lambda} = \frac{2 \pi \alpha}{\lambda^2} \left(\frac{g_D n_P}{e} \right) ^2 \left(1-\frac{1}{\beta^2 n_P^2} \right)
\end{equation}
\end{linenomath}
Thus, a minimally charged mono\-pole generates $(g_D n_P/e)^2\approx8200$  
times more Cherenkov radiation 
in ice 
compared to an electrically charged particle with the same velocity. This is shown in Fig.  \ref{num_photons}.
 
In addition to this effect, a (mildly) relativistic mono\-pole knocks electrons off their binding with an atom. 
These high-energy $\delta$-electrons can have velocities above the Cherenkov threshold.
For the production of $\delta$-electrons the differential cross-section of Kasama, Yang and Goldhaber (KYG) is used that allows to calculate the energy transfer of the mono\-pole to the 
$\delta$-electrons and therefore the resulting output of indirect Cherenkov light \cite{Wu76, KYG77}. 
The KYG cross section was calculated using QED, particularly dealing with the monopole's vector potential and its singularity \cite{Wu76}.
Cross sections derived prior to KYG, such as the so-called Mott cross section \cite{Bauer51, Cole51, Ahlen75}, are only semi-classical approximations because the mathematical tools had not been developed by then. Thus, in this work the state-of-the-art KYG cross section is used to derive the light yield. 
The number of photons derived with the KYG 
and Mott 
cross section are shown in Fig. \ref{num_photons}. 
Above the Cherenkov threshold indirect Cherenkov light is negligible for the total light yield.

Using the KYG cross section the energy loss of magnetic mono\-poles per unit path length $dE/dx$  can be calculated \cite{Ahlen78}
\begin{linenomath} 
\begin{equation}
\begin{aligned}
	\frac{dE}{dx}=  \frac{4\pi N_e g_D^2 e^2}{m_e c^2}  & \left[ \ln{\frac{2 m_e c^2 \beta^2 \gamma^2}{I}}+\frac{K(g_D)}{2} \right. \\
																					 & \left. -\frac{\delta+1}{2}-B(g_D)  \right]
	\label{eq:Energyloss}
\end{aligned}
\end{equation}
\end{linenomath}
where $N_e$ is the electron density, $m_e$ is the electron mass, $\gamma$ is the Lorentz factor of the mono\-pole, $I$ is the mean ionization potential, $K(g_D)$ is the QED correction derived from the KYG cross section, $B(g_D)$ is the Bloch correction and $\delta$ is the density-effect correction \cite{Sternheimer84}.

Luminescence  is the third process which may be  considered in the velocity range. 
It has been shown that pure ice exposed to ionizing radiation emits luminescence light \cite{Grossweiner52, Grossweiner54}. The measured time distribution of luminescence light is fit well by  several overlapping decay times which hints at several different excitation and de-excitation mechanisms \cite{Quickenden82}. The most prominent wavelength peaks are within the DOM acceptance of about $300\,\textrm{nm}$ to $600\,\textrm{nm}$ \cite{Quickenden82, Spice13}.
The mechanisms are highly dependent on temperature and ice structure. Extrapolating the latest measurements  of luminescence light $dN_{\gamma}/dE$ \cite{Quickenden82, Baikal08}, the brightness $ dN_\gamma / dx $
\begin{linenomath}
\begin{equation}
\frac{dN_\gamma}{dx}=\frac{dN_\gamma}{dE}  \cdot \frac{dE}{dx}
\end{equation}
\end{linenomath} 
could be at the edge of IceCube's sensitivity where the energy loss is calculated with Eq. \ref{eq:Energyloss}.
This means that it would not be dominant 
above $0.5\, c$.
The resulting brightness is almost constant for a wide velocity range from $0.1\,c$ to $0.95\, c$. 
Depending on the actual brightness, luminescence light could be a promising method to detect mono\-poles with lower velocities.
Since measurements of $dN_\gamma/dE$ are still to be done  for the parameters given in IceCube, luminescence has to be neglected in the presented analyses 
which is a conservative approach leading to lower limits.


\section{ Simulation \label{sec:Sim}}

The simulation of an IceCube event comprises several steps. 
First, a particle is generated, i.e. given its start position, direction and velocity. Then it is propagated, taking into account decay and interaction probabilities, and propagating all secondary particles as well. 
When the particle is close to the detector, the Cherenkov light is generated and the photons are propagated through the ice accounting for its properties. Finally the response of the PMT and DOM electronics is simulated including the generation of noise and the triggering and filtering of an event (see Sec. \ref{sec:IC}). 
From the photon propagation onwards, the simulation is handled identically for background and monopole signal. 
However the photon propagation is treated differently in the two analyses presented below 
due to improved ice description and photon propagation software available for the latter analysis.

\subsection{Background generation and propagation}

The background of a mono\-pole search consists of all other known particles which are detectable by IceCube. 
The most abundant background are muons or muon bundles produced in air showers caused by cosmic rays. These were modeled using the cosmic ray models Polygonato \cite{Hoerandel03} for the highly relativistic and GaisserH3a \cite{Gaisser11} for the mildly relativistic analysis. 

The majority of neutrino induced events are caused by neutrinos created in the atmosphere. Conventional at\-mo\-spher\-ic neutrinos, produced by the decay of charged pions and kaons, are dominating the neutrino rate from the GeV to the TeV range \cite{Honda06}. Prompt neutrinos, which originate from the decay of heavier mesons, i.e. containing a charm quark, are strongly suppressed at these energies \cite{SarcevicStd08}.

Astrophysical neutrinos, which are the primary objective of IceCube, have only recently been found \cite{HESE13, HESE3}. 
For this reason they are only taken into account as a background in the mildly relativistic analysis, using the fit result for the astrophysical flux from Ref. \cite{HESE3}. 

Coincidences of all background signatures are also taken into account.

\begin{figure}
	\includegraphics[width=3in]{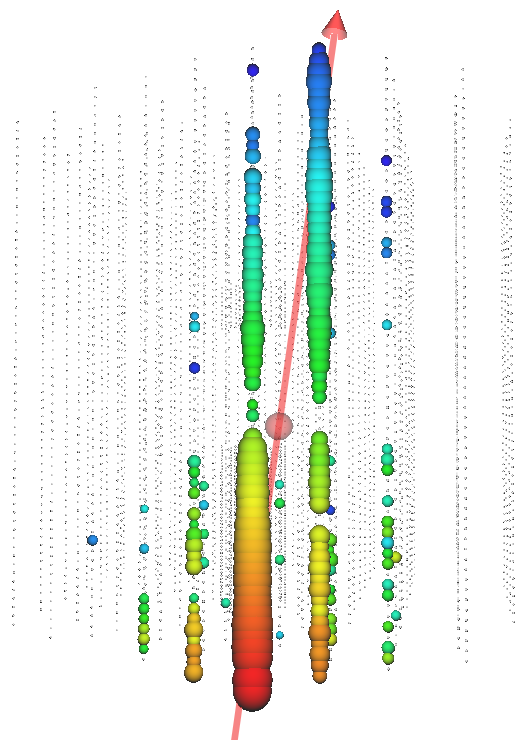}
	\caption{
		Event view of a simulated 
		magnetic mono\-pole with a velocity of $0.83\, c$ using both direct and indirect Cherenkov light. The monopole track is created with a zenith angle of about $170^{\circ}$ in upward direction. 
		The position of the IceCube DOMs are shown with gray spheres.
		Hit DOMs are visualized with colored spheres. Their size is scaled with the number of recorded photons. 
		The color denotes the time development from red to blue. The red line shows the reconstructed track 
		which agrees with the true direction
		} 
	\label{fig:Sim2}
\end{figure}

\subsection{Signal generation and propagation}

Since the theoretical mass range for magnetic mono\-poles is broad (see Sec. \ref{sec:Intro}), and the Cherenkov emission is independent of the mass, signal simulation is focused simply on a benchmark mono\-pole mass of $10^{11} \; \textrm{GeV}$ without limiting generality. Just the ability to reach the detector after passing through the Earth depends on the mass predicted by a mono\-pole model.
The parameter range for mono\-poles producing a recordable light emission inside IceCube is governed by the velocities needed to produce (indirect) Cherenkov light. 

The starting points of the simulated mono\-pole tracks are generated uniformly distributed around the center of the completed detector and pointing towards the detector.
For the highly relativistic analysis  the simulation could be run at specific monopole velocities only and so the characteristic velocities $0.76\, c$, $0.8\, c$, $0.9\, c$ and $0.995\, c$, were chosen. 

Due to new software, described in the next sub-section, in the simulation for the mildly relativistic analysis the mono\-poles can be given an arbitrary characteristic velocity $v$ below $0.99\, c$. The light yield from indirect Cherenkov light fades out below $0.5\, c$. To account for the smallest detectable velocities the lower velocity limit was set to $0.4\, c$ in simulation.

The simulation also accounts for monopole deceleration via energy loss.
This information is needed to simulate the light output.

\subsection{Light propagation}

In the highly relativistic analysis the photons from direct Cherenkov light are propagated using Photonics \cite{Photonics07}.
A more recent and GPU-enabled software propagating light in IceCube is PPC \cite{Spice13} which is used in the mildly relativistic analysis. 
The generation of direct Cherenkov light, following Eq. \ref{eq:DN}, was implemented into PPC in addition to the variable Cherenkov cone angle (Eq. \ref{EQ:CherCone}). For indirect Cherenkov light a parametrization of the distribution in Fig. \ref{num_photons} is used.

Both simulation procedures are consistent with each other and deliver a signal with the following topology: 
through-going tracks, originating from all directions,  with constant velocities and brightness inside the detector volume, see Fig. \ref{fig:Sim2}. All these properties are used to discriminate the monopole signal from the background in IceCube.


\section{Highly relativistic analysis \label{sec:IC40}}

This analysis covers the velocities above the Cherenkov threshold $v_C\approx0.76\,c$ and it is based on the IC40 data recorded from May 2008 to May 2009. This comprises about 346 days of live-time or 316 days without the burn sample. 
The live-time is the recording time for clean data. 
The analysis for the IC40 data follows the same conceptual design as a previous analysis developed for the IC22 data \cite{Christy13},
focusing on a simple and easy to interpret set of variables. 

\subsection{Reconstruction}

The highly relativistic analysis uses spatial and
timing information from the following sources: all DOMs, fulfilling the next or next-to-nearest neighbor condition (described in section \ref{sec:IC}), and DOMs that fall into the topmost 10\% of the collected-charge distribution for that event which are supposed to record less scattered photons. This selection allows definition of variables that benefit from either large statistics or precise
timing information.

\begin{figure}
	
	\includegraphics[width=2.5in]{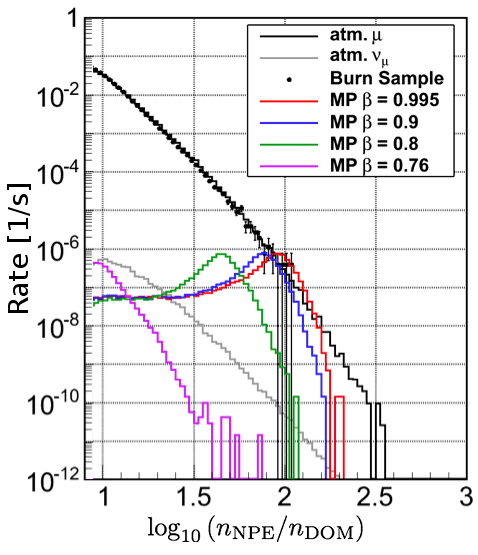}
	\centering
	\caption{The relative brightness after the first two cuts on $n_{\textrm{DOM}}$ and $n_{\textrm{NPE}}/n_{\textrm{DOM}}$. 
		The expected distributions from mono\-poles (MP) of different velocities is shown for comparison
		} 

	\label{fig:L0_IC40}
\end{figure}

\subsection{Event selection}

The IC40 analysis selects events based on their relative brightness, arrival direction, and
velocity. Some additional variables are used to identify and reject events with poor track reconstruction
quality. The relative brightness is defined as the average number of photo-electrons per
DOM contributing to the event. This variable has more dynamic range compared with the number of hit DOMs. 
The distribution of this variable after applying the first two quality cuts, described in Tab. \ref{tab:cutsIC40}, is shown in Fig. \ref{fig:L0_IC40}.
Each event selection step up to the final level is optimized to minimize the
background passing rate while keeping high 
signal efficiency, see Tab. \ref{tab:cutsIC40}.

\begin{figure}
	\includegraphics[width=3.1in]{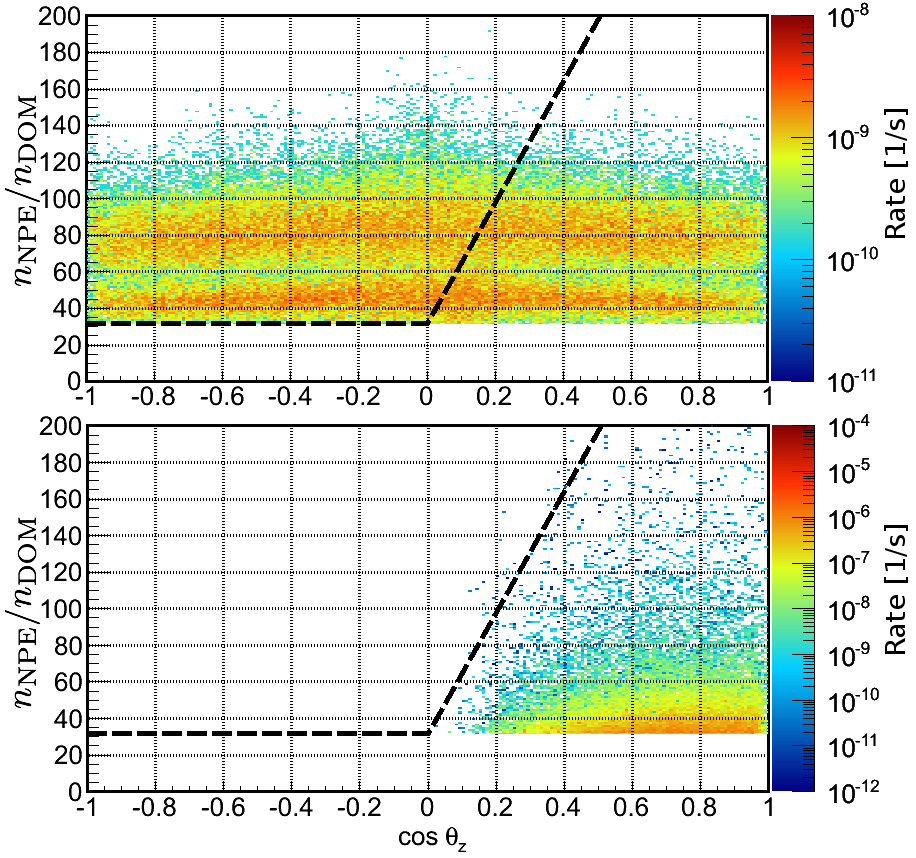}
	\centering
	\caption{
		Comparison of signal distribution (top) vs. at\-mo\-spher\-ic muon background (bottom) for the final cut. The signal is the composed out of the sum of monopoles with $\beta = 0.995, 0.9, 0.8$
	} 
	\label{fig:LastCut_IC40}
\end{figure}

The final event selection level aims to remove the bulk of the remaining background, mostly 
consisting of downward going at\-mo\-spher\-ic muon bundles. However, the dataset is first split in two
mutually exclusive subsets with low and high brightness.
This is done in order to isolate a well known discrepancy between experimental and simulated data in the direction distribution near the horizon which is caused by deficiencies in simulating air shower muons at high inclinations \cite{Posselt13}.

Since attenuation is stronger at large zenith angles $\theta_z$, the brightness of the resulting events is reduced and the discrepancy is dominantly located in the low brightness subset. Only simulated mono\-poles with $v = 0.76\;c$ significantly populate this subset. The
final selection criterion for the low brightness subset is $\cos \theta_z < -0.2$ where $\theta_z$
is the reconstructed arrival angle with respect to the zenith. For the high brightness subset
a 2-dimensional selection criterion is used as shown in Fig. \ref{fig:LastCut_IC40}. The two variables
are the relative brightness described above and the cosine of the arrival angle. Above the
horizon ($\cos \theta_z > 0$), where most of the background is located, the selection threshold
increases linearly with increasing $\cos \theta_z$. Below the horizon the selection has no
directional dependence and values of both ranges coincide at $\cos \theta_z = 0$. The
optimization method applied here is the  model rejection potential (MRP) method 
described in \cite{Christy13}.

\begin{table*}
	
	\caption{Uncertainties in both analyses.  
		For the mildly relativistic analysis the average for the whole velocity range is shown. See Fig. \ref{fig:Vel} for the velocity dependence
		}
	\label{tab:uncertaintiesIC40}
	\label{tab:uncertaintiesIC86}    
	
	\begin{tabular}{p{2.2cm}|p{2.1cm}p{2cm}p{1.4cm}p{1cm}p{1cm}p{1.2cm}|p{1cm}p{1.4cm}}
		\hline\noalign{\smallskip}
		Conf. & IC40 &&&&&& IC86  &\\
		\noalign{\smallskip}\hline\noalign{\smallskip}
		Type  & Atm. $\nu_\mu$  in \%  &  & Signal in \%  & & &  & $\nu_{\mu}$ in \% & Signal in \% \\
		&         high $n_{\textrm{NPE}}/n_{\textrm{DOM}}$ & low $n_{\textrm{NPE}}/n_{\textrm{DOM}}$   & $\beta=0.995$& $\beta=0.9$ & $\beta=0.8$& $\beta=0.76$ & \multicolumn{2}{c}{$0.4 \leq \beta \leq 0.99$}\\
		\noalign{\smallskip}\hline\noalign{\smallskip}
		
		Statistics								& $\,\,\,\,3.7$  & $\,\,\,\,6.4$ & 0.7 & 0.7& 0.8 & 0.5  & $\,\,\,\,6.8$ 				& 0.4	\\ 
		DOM Efficiency  & 25.9 & 40.8  & 3.2& 2.7& 5.3 & 15.6 & $\,\,\,\,8.1$ 				& 1.3 	\\ 
		Light Propagation     		     & 20.5 & 34.9 & 2.9& 2.4& 3.6 & 6.1 & 12.4							& 2.7 \\
		Flux 
		& 25.8 & 26.1 & - & - & - & - & $\,\,\,\,8.2$			    & - \\ 
		Re-sam\-pling & - & - & - & - & - & - & see text 		 & see text		\\
		\noalign{\smallskip}\hline\noalign{\smallskip}
		
		Total 								& 42.1 & 60.0 & 4.4 & 3.7 & 6.4 & 16.7 & 16.9					& 3.0\\
	
		\noalign{\smallskip}\hline
	\end{tabular}
\end{table*}

\subsection{Uncertainties and Flux Calculation\label{sec:uncertainty}}

Analogous to the optimization of the final event selection level, limits on the mono\-pole flux are
calculated using a MRP method. 
Due to the blind approach of the analysis these are derived from Monte Carlo simulations,
which contain three types of uncertainties: (1) Theoretical uncertainties in the simulated models,
(2) Uncertainties in the detector response, and (3) Statistical uncertainties. 

For a given mono\-pole-velocity the limit then follows from
\begin{linenomath}
\begin{equation}
  \Phi_{\alpha} = \mathrm{MRP} \cdot \Phi_0 = \frac{\bar{\mu}_{\alpha}(n_{\mathrm{obs}})}{\bar{n}_{\mathrm{s}}} \Phi_0 
\end{equation}
\end{linenomath}
where $\bar{\mu}_{\alpha}$ is an average Feldman-Cousins (FC) upper limit with confidence $\alpha$, 
which depends on the number of observed events $n_{\mathrm{obs}}$. Similarly, though derived
from simulation, $\bar{n}_{\mathrm{s}}$ is the average expected number of observed signal events assuming
a flux $\Phi_0$ of magnetic mono\-poles. Since $\bar{n}_{\mathrm{s}}$ is proportional to $\Phi_0$
the final result is independent of whichever initial flux is chosen.

The averages can be independently expressed as weighted sums over values of
$\mu_{\alpha}(n_{\mathrm{obs}}, n_{\mathrm{bg}})$ and $n_{\mathrm{s}}$ respectively with the FC
upper limit here also depending on 
the number of expected background events 
$n_{\mathrm{bg}}$ 
obtained from simulation. 
The weights are then the probabilities for
observing a particular value for $n_{\mathrm{bg}}$ or $n_{\mathrm{s}}$. In the absence of
uncertainties this probability has a Poisson distribution with the mean set to the expected number of
events $\lambda$ derived from simulations. 
However, in order to extend the FC approach to account for uncertainties, the distribution
\begin{linenomath}
\begin{equation}
  \label{Eq:GeneralPDF}
	\mathrm{PDF}(n|\lambda,\sigma) = \int \frac{(\lambda+x)^{n}\;e^{-\lambda-x}}{n!} \cdot w(x|\sigma) \;dx
\end{equation}
\end{linenomath}
is used instead to derive $n_{\mathrm{bg}}$ and $n_{\mathrm{s}}$.
This is the weighted average of Poisson distributions where the mean
value varies around the central value $\lambda$ and the variance $\sigma^2$ is the quadratic
sum of all individual uncertainties. Under the assumption that individual contributions to the
uncertainty are symmetric and independent, the weighting function \(w(x|\sigma)\) is a normal
distribution with mean 0 and variance \(\sigma^2\). However, the Poisson distribution is only
defined for positive mean values. Therefore a truncated normal distribution with the boundaries
\(-\lambda\) and \(+\infty\) is used as the weighting function instead. 


\section{Mildly relativistic analysis \label{sec:IC86}}

This analysis uses on the data recorded from May 2011 to May 2012. It comprises about 342 days (311 days without the burn sample) of live-time. The signal simulation covers the velocity range of $0.4\,c$ to $0.99\, c$. The optimization of cuts and  machine learning is done on a limited velocity range  $<\,0.76\, c$ to focus on lower velocities where indirect Cherenkov light dominates.

\subsection{Reconstruction}

Following the filters, described in Sec. \ref{sec:IC}, further processing of the events is done by splitting coincident events into sub-events using a time-clustering algorithm. 
This is useful to reject hits caused by PMT after-pulses which appear several microseconds later than signal hits.

For quality reasons events are required to have 6 DOMs on 2 strings hit, see Tab. \ref{tab:cutsIC86}. 
The remaining events are handled as tracks reconstructed with 
an improved version \cite{ImprLineFit14} of the LineFit algorithm, mentioned in Sec. \ref{sec:IC}. 
Since the main background  in IceCube are muons from air showers which cause a down-going track signature, a cut on the reconstructed zenith angle below $86^{\circ}$ removes most of this background.

\begin{figure}
	
	\includegraphics[width=3in]{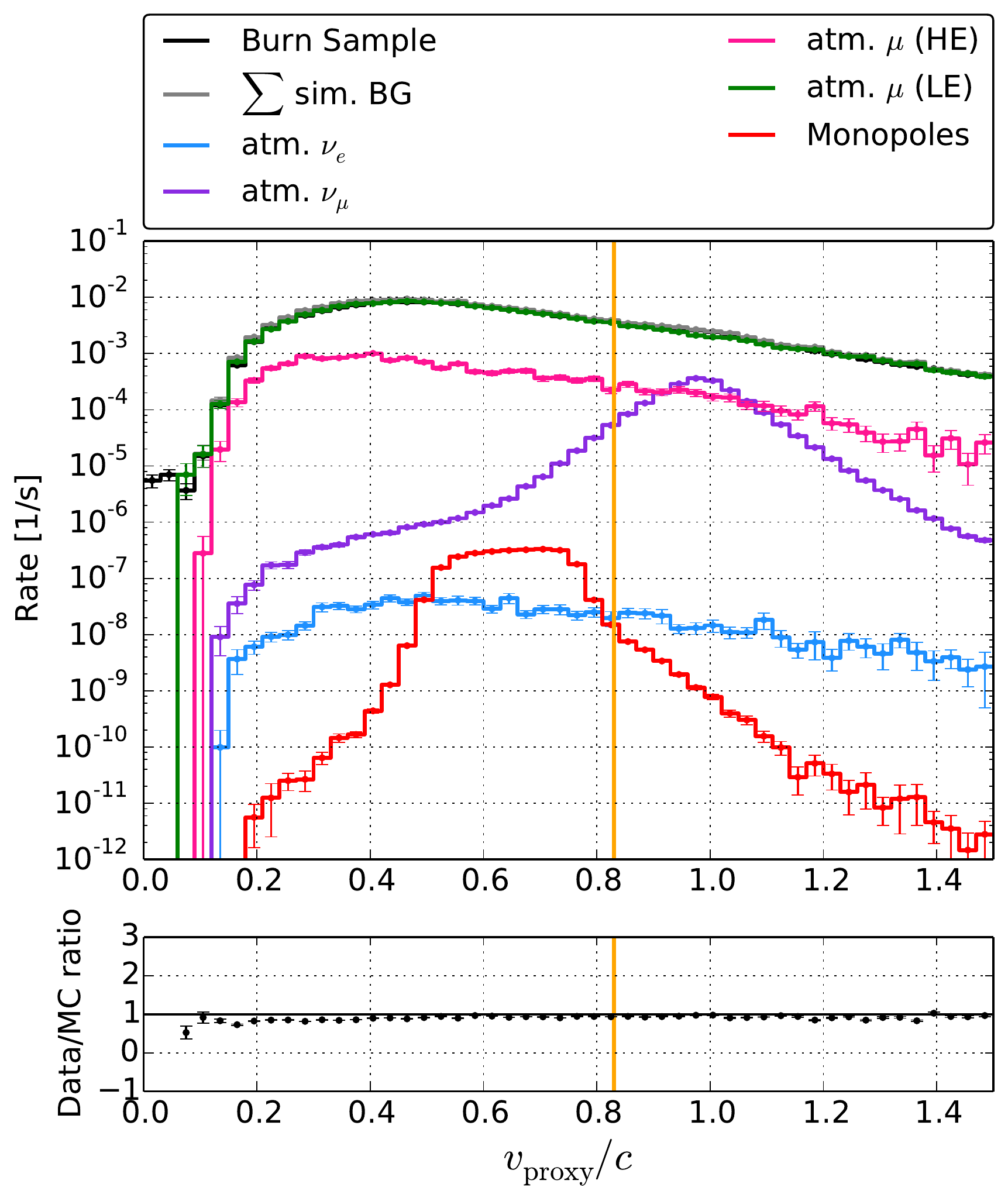} 
	\centering
	\caption{Estimated velocity after event reconstruction. In this plot only mono\-poles with a simulated true velocity below $0.76\, c$ are shown and a cut on the reconstructed velocity at $0.83\, c$. These restrictions were only used for training to focus on this range and released for sensitivity calculation and unblinding. 
	Superluminal velocity values occur because of the simplicity of the chosen reconstruction algorithm which may lead 
	to mis-reconstructed events that can be discarded. 
	The air shower background is divided into high (HE) and low energy (LE) primary particle energy at $100\,\textrm{TeV}$. The recorded signals differ significantly and are therefore treated with different variables and cuts
	}
	\label{fig:LF_Cut}
\end{figure}

Figure \ref{fig:LF_Cut} shows the reconstructed particle velocity at this level.
The rate for at\-mo\-spher\-ic muon events has its maximum at low velocities. This is due to mostly coincident events remaining in this sample.
The muon neutrino event rate consists mainly of track-like signatures and is peaked at the velocity of light. 
Dim events or events traversing only part of the detector are reconstructed with lower velocities which leads to the smearing of the peak rate for muon neutrinos and mono\-pole simulations. Electron neutrinos usually produce a cascade of particles (and light) when interacting which is easy to separate from a track signature. The velocity reconstruction for these events results mainly in low velocities which can also be used for separation from signal. 

\subsection{Event selection}

In contrast to the highly relativistic analysis, machine learning was used. A boosted decision tree (BDT) \cite{BDT95} was chosen to account for limited background statistics. The multivariate method was embedded in a re-sampling method. 
This was combined with additional cuts to reduce the background rate and prepare the samples for an optimal training result.
Besides that, these straight cuts reduce cascades, coincident events, events consisting of pure noise, improve reconstruction quality, and remove short tracks which hit the detector at the edges. See a list of all cuts in Tab. \ref{tab:cutsIC86}. To train the BDT on lower velocities an additional cut on the maximal velocity $0.82\, c$ is used only during training which is shown in Fig. \ref{fig:LF_Cut}. Finally a cut on the  penetration depth of a track, measured from the bottom of the detector, is performed. 
This is done to lead the BDT training  to a suppression of air shower events underneath the neutrino rate near the signal region, as can  be seen in Fig. \ref{fig:pull2}.

Out of a the large number of variables provided by standard and mono\-pole reconstructions 15 variables  were chosen for the BDT using a tool called mRMR (Minimum Redundancy Maximum Relevance) \cite{MRMR05}. These 15 variables are described in Tab. \ref{tab:bdtIC86}. 
With regard to the next step it was important to choose variables which show a good data - simulation agreement so that the BDT would not be trained on unknown differences between simulation and recorded data. 
The resulting BDT score distribution in Fig. \ref{fig:pull1} shows a good signal vs. background separation with reasonable simulation - data agreement. The rate of at\-mo\-spher\-ic muons and electron neutrinos induced events is suppressed sufficiently compared to the muon neutrino rate near the signal region. The main background is muon neutrinos from air showers.

\subsection{Background Expectation}

\begin{figure}
	\centering
	\includegraphics[height=3.5in]{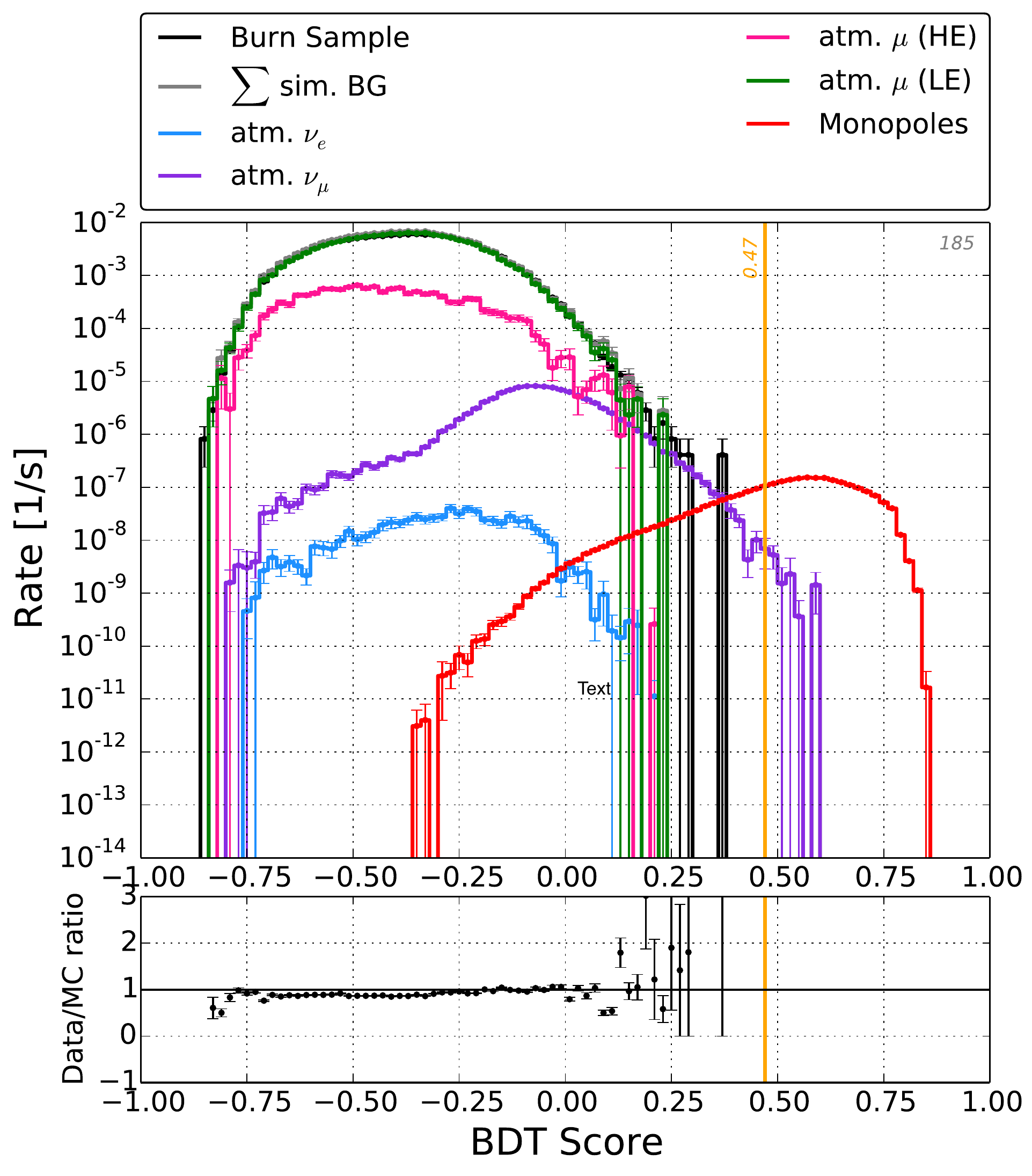}
	
	\caption{Distribution of one BDT trained on 10\% of the burn sample. 
		The cut value which is chosen using Fig. \ref{fig:pull2} is shown with the orange line. Statistical errors per bin are drawn
	} 
	\label{fig:pull1}
\end{figure}

To calculate the background expectation a method inspired by bootstrapping is used \cite{Bootstrap79}, called pull-validation \cite{Luenemann2015}. 
Bootstrapping is usually used to smooth a distribution by resampling the limited available statistics. Here, the goal is to smooth especially the tail near the signal region in Fig. \ref{fig:pull1}.

Usually 50\% of the available data is chosen to train a BDT which is done here just for the signal simulation. Then the other 50\% is used for testing. Here, 10\% of the burn sample are chosen randomly, to be able to consider the variability in the tails of the background. 
 
\begin{figure}
	
	\includegraphics[height=3.5in]{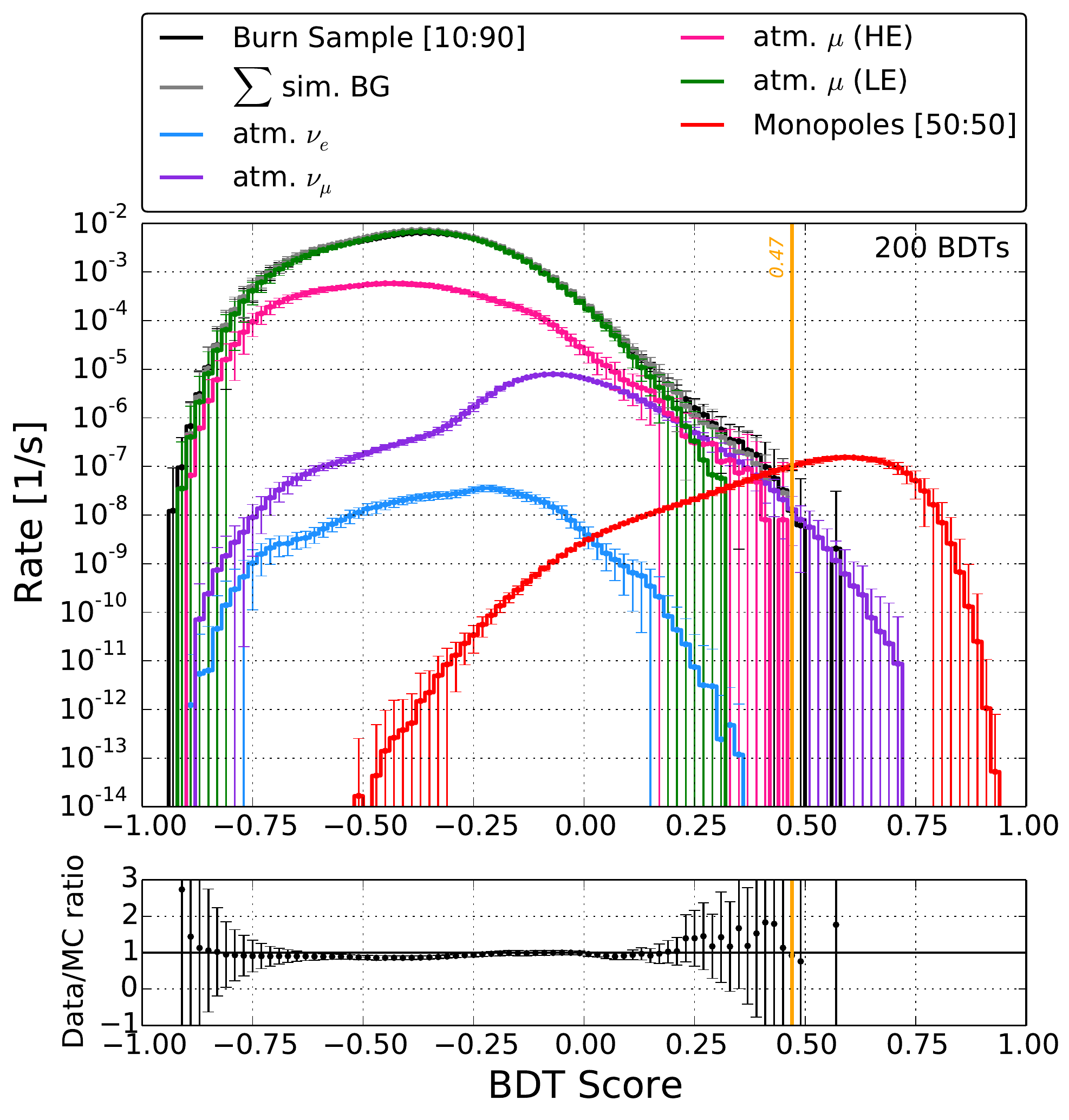}
	\centering
	\caption{
		Average of 200 BDTs. An example of one contributing BDT is shown in Fig. \ref{fig:pull1}. 
		In each bin the mean bin height in 200 BDTs is shown with the standard deviation as error bar.  
		Based on this distribution the MRF is calculated and minimized to choose the cut value
	} 
	\label{fig:pull2}
\end{figure}

Testing the BDT on the other 90\% of the burn sample leads to an extrapolation of the tail into the signal region. 
This re-sam\-pling and BDT training / testing is repeated 200 times,  each time choosing a random 10\% sample. 
In Fig. \ref{fig:pull2} the bin-wise average and standard deviation of 200 BDT score distributions are shown.

By BDT testing, 200 different BDT scores are assigned to each single event. The event is then transformed into a  probability density distribution. 
When cutting on the BDT score distribution in Fig. \ref{fig:pull2} a single event $i$ is neither completely discarded nor kept, but it is kept with a certain probability $p_i$ which is calculated as a weight. 
The event is then weighted in total with $W_i=p_i \cdot w_i$ using its survival probability and the weight $w_i$ from the chosen flux spectrum. 
Therefore, many more events contribute to the cut region compared to a single BDT which reduces the uncertainty of the background expectation.

To keep the error of this statistical method low, the cut on the averaged BDT score distribution is chosen near the value where statistics in a single BDT score distribution vanishes.

The developed re-sam\-pling method gives the expected background rate including an uncertainty for each of the single BDTs. Therefore one BDT was chosen randomly for the unblinding of the data.

\subsection{Uncertainties}

The uncertainties of the re-sam\-pling method were investigated thoroughly.
The Poissonian error per bin is negligible because of the averaging of 200 BDTs. Instead, there are 370 partially remaining events which contribute to the statistical error. 
This uncertainty $\Delta_{\textrm{contr}}$ is estimated by considering the effect of omitting individual events $i$ of the 370 events from statistics
\begin{linenomath}
\begin{equation}
\Delta_{\textrm{contr}} = \max_i \left( \frac{w_i p_i}{\sum_i w_i p_i} \right)
\end{equation}
\end{linenomath}
 Datasets with different simulation parameters for the detector properties are used to calculate the according uncertainties. The values of all calculated uncertainties are shown in Tab. \ref{tab:uncertaintiesIC86}.

The robustness of the re-sam\-pling method was verified additionally by varying all parameters and cut values of the analysis. 
Several fake unblindings were done by training the analysis on a 10\% sample of the burn sample, optimizing the last cut and then applying this event selection on the other 90\% of the burn sample. This proves reliability by showing that the previously calculated background expectation is actually received with increase of statistics by one order of magnitude. 
The results were mostly near the mean neutrino rate, only few attempts gave a higher rate, but no attempt exceeded the calculated confidence interval. 

\begin{figure*}
	\includegraphics[width=6.5in]{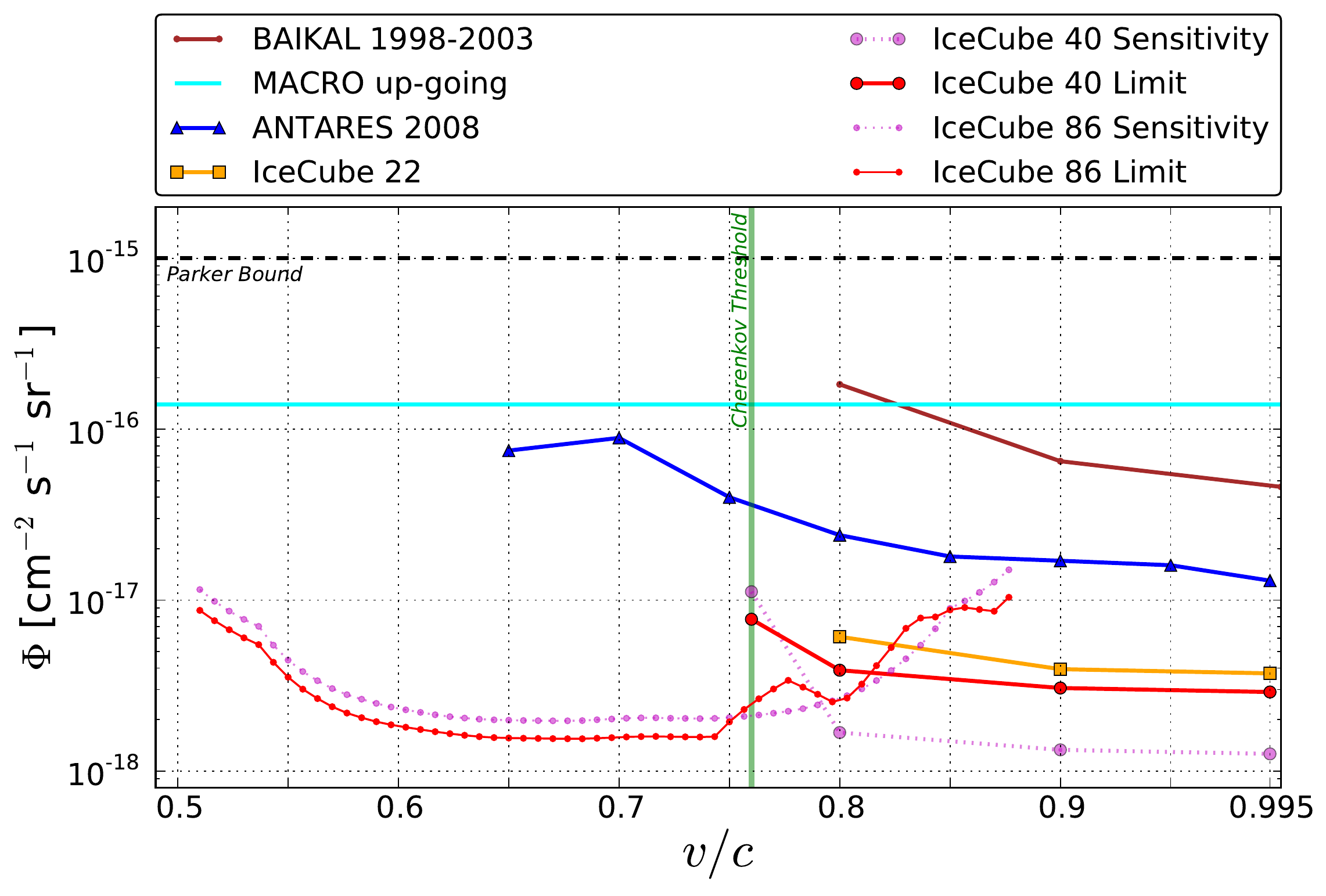}
	\centering
	
	\caption{Sensitivities (magenta) and final limits (red) of both analysis at certain characteristic velocities compared to other limits.  The lines are only drawn to guide the eyes. 
		Other limits are from BAIKAL \cite{Baikal08}, 
		ANT\-ARES \cite{Antares12}, IceCube 22 \cite{Christy13}, MACRO \cite{Macro02}. Also shown is the Parker limit described in the text \cite{Parker70}
	} 	
	\label{fig:limit}
\end{figure*}

The rate of the background events has a variability in all 200 BDTs of up to 5 times the mean value of 0.55 events per live-time (311 days) when applying the final cut on the BDT score. This contribution is dominating the total uncertainties. Therefore not a normal distribution but the real distribution is used for further calculations.
This distribution is used as a probability mass function in an extended Feldman Cousin approach to calculate the 90\% confidence interval, as described in Sec. \ref{sec:uncertainty}. The final cut at BDT score 0.47 is chosen near the minimum of the model rejection factor (MRF) \cite{FeldmanCousins98}. To reduce the influence of uncertainties it was shifted to a slightly lower value. The sensitivity for many different velocities is  calculated as described in Sec. \ref{sec:uncertainty} and shown in Fig. \ref{fig:limit}.  This gives an 90\% confidence upper limit of  3.61 background events. 
The improvement of sensitivity compared to recent limits by ANT\-ARES \cite{Antares12} and MACRO \cite{Macro02} reaches from one to almost two orders of magnitude which reflects a huge detection potential.

\section{Results \label{sec:result}}

After optimizing the two analyses on the burn samples, the event selection was adhered to 
and the remaining 90\% of the experimental data were processed ("unblinded").
The corresponding burn samples were not included while calculating the final limits.

\subsection{ Result of the highly relativistic analysis}

In the analysis based on the IC40 detector configuration three events remain, one in the low
brightness subset and two in the high brightness subset. The low brightness event is consistent
with a back\-ground-
only observation with 2.2 expected background events. 
The event itself shows characteristics typical for a neutrino induced muon. For the high brightness subset, with
an expected background of $0.1$ events, the observation of two events apparently contradicts the back\-ground-only hypothesis. 
However, a closer analysis of the two events reveals that they are unlikely to be caused by mono\-poles. 
These very bright events do not have a track like signature but a spheric development only partly contained in the detector. 
A possible explanation is  the now established flux of cosmic neutrinos which was not included in
the background expectation for this analysis. 
IceCube's unblinding policy prevents any claims on these events or reanalysis with changed cuts as have been employed with IC22 \cite{Christy13}.
Instead they are treated as an upward fluctuation of the background weakening the limit.
The final limits outperform previous limits and are shown in Tab. \ref{tab:limits} and Fig. \ref{fig:limit}.
These limits can also be used as a conservative limit for  $v>0.995\,c$ without optimization for high values of Lorentz factor $\gamma$ as the expected monopole signal is even brighter due to stochastic energy losses which are not considered here.

\begin{figure}
	
	\includegraphics[width=3.2in]{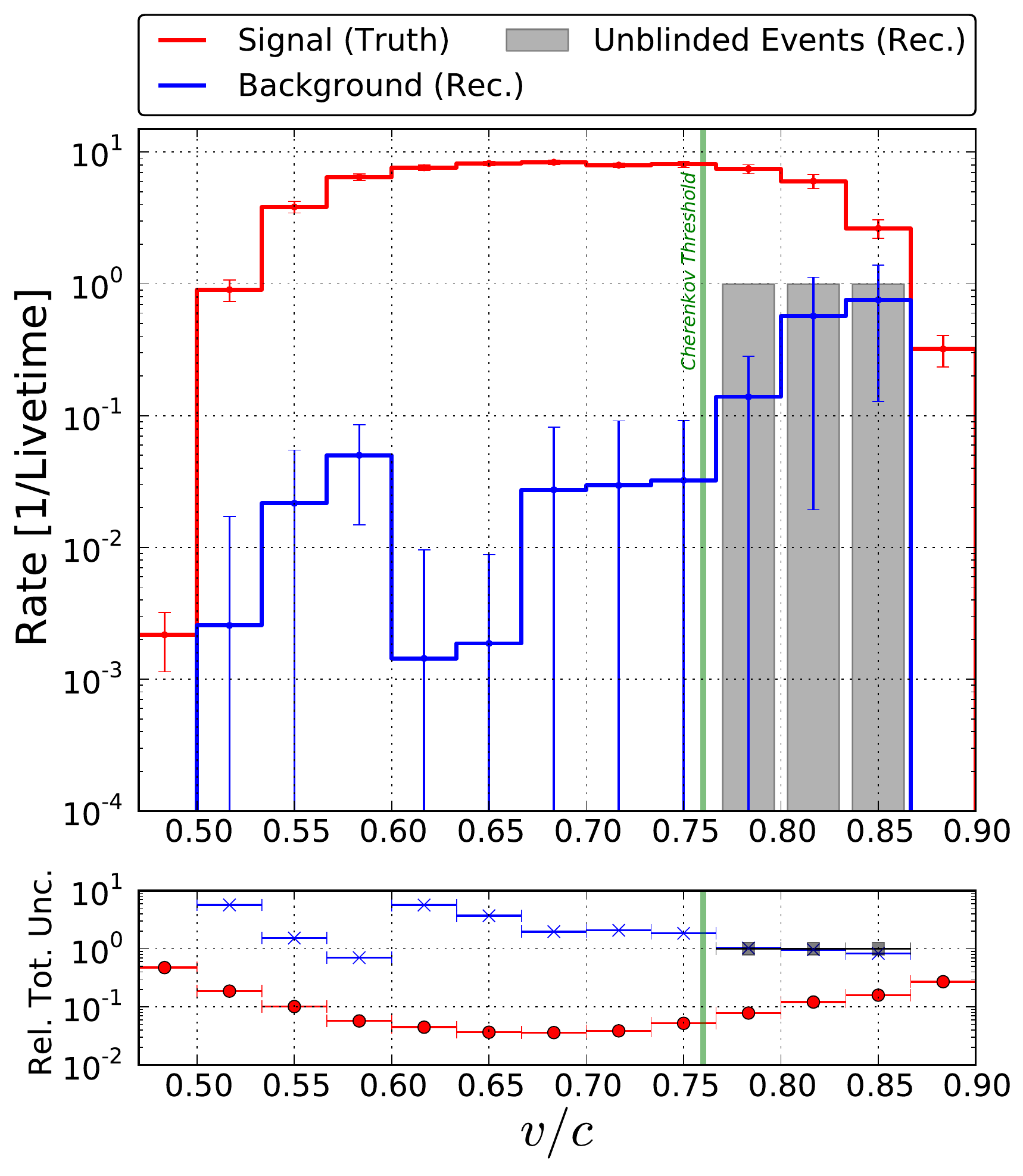}
	\centering
	\caption{Signal and background rates per characteristic monopole velocity which are used to calculate the final limits. 
		Reconstructed velocity is used for background and true simulated velocity for signal. 
		The lower part of the plot shows the velocity dependence of the uncertainties including the re-sam\-pling uncertainty which dominates. 
		The different contributions to the uncertainties are listed in Tab. \ref{tab:uncertaintiesIC86}
	}
	\label{fig:Vel}
\end{figure}

\begin{figure}
	
	\includegraphics[width=2.8in]{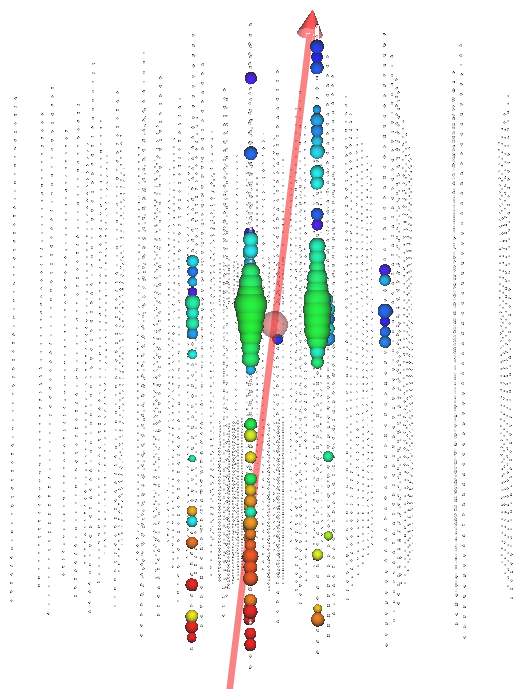}
	\caption{One of the three events which were selected in the mildly relativistic analysis with a BDT Score of 0.53. The reconstructed parameters of this event are the same as in Fig. \ref{fig:Sim2}. 
		In this event, 110 DOMs were hit on 8 strings.
		It has a brightness of $595 \; \textrm{NPE}$  
		and causes an after-pulse.
		The position of the IceCube DOMs are shown with small gray spheres.
		Hit DOMs are visualized with colored spheres. Their size is scaled with the brightness of the hit. The color denotes the time development from red to blue. The red line shows the reconstructed track} 
	\label{fig:MMM3}
\end{figure}

\subsection{Result of the mildly relativistic analysis}

In the mildly relativistic analysis three events remain after 
all cuts which is within the confidence interval of up to 3.6 events and therefore consistent with a background only observation. All events have reconstructed velocities above the training region of $ 0.76c $. This is compared to the expectation from simulation in Fig. \ref{fig:Vel}. 
Two of the events show a signature which is clearly incompatible with a mono\-pole signature when investigated by eye because they are stopping within the detector volume. The third event, shown in Fig. \ref{fig:MMM3}, may have a mis-reconstructed velocity due to the large string spacing of IceCube.  
However, its signature is comparable with a mono\-pole signature with a reduced light yield 
than described in Sec. \ref{sec:Cher}. 
According to simulations, a monopole of this reconstructed velocity would emit about 6 times the observed light. 

To be comparable to the other limits shown in Fig. \ref{fig:limit} the final result of this analysis is calculated for different characteristic mono\-pole velocities at the detector. 
The bin width of the velocity distribution in Fig. \ref{fig:Vel} is chosen to reflect 
the error on the velocity reconstruction. Then, the limit in each bin is calculated and normalized which gives a 
step function. To avoid the bias on a histogram by choosing different histogram origins, five different starting points are chosen for the distribution in  Fig. \ref{fig:Vel} and the final step functions are averaged  \cite{Haerdle07}.

The final limit is shown in Fig. \ref{fig:limit} and Tab. \ref{tab:limits}
together with the limits from the highly relativistic analysis and other recent limits.


\section{Discussion \label{sec:End}}

The resulting limits are placed into context by considering indirect theoretical limits and previous experimental results.
The flux $\Phi$ of magnetic mono\-poles can be constrained model independently by astrophysical arguments to  $\Phi_{\textrm{P}} \leq 10^{-15} \; \allowbreak \textrm{cm}^{-2} \textrm{s}^{-1} \textrm{sr}^{-1}$ for a monopole mass below $10^{17} \; \allowbreak \textrm{GeV}/c^2$. This value is the so-called Parker bound~\cite{Parker70} which has already been surpassed by several experiments as shown in Fig. \ref{fig:limit}. The most comprehensive search for mono\-poles, regarding the velocity range, was done by the MACRO collaboration using different detection methods \cite{Macro02}. 

More stringent flux limits have been imposed by using larger detector volumes, provided by high-energy neutrino telescopes, such as ANT\-ARES \cite{Antares12}, BAI\-KAL \cite{Baikal08}, AMAN\-DA \cite{Amanda10},  and IceCube \cite{Christy13}. 
The current best limits for  non-rel\-a\-tivistic velocities ($\leq 0.1\,c$) have been established  by IceCube, constraining the flux down to a level of $\Phi_{\textrm{90\%}}\geq10^{-18} \; \allowbreak \textrm{cm}^{-2} \textrm{s}^{-1} \textrm{sr}^{-1}$ \cite{Schoenen14}. These limits hold for the proposal that monopoles catalyze proton decay. 
The analysis by ANT\-ARES is the only one covering the mildly relativistic velocity range ($\geq 0.625\,c$) using a neutrino detector, to date.
However, using the KYG cross section for the $\delta$-electron production would extend their limits to lower velocities. 
The Baksan collaboration has also produced limits on a monopole flux \cite{Baksan06}, both at slow and relativistic velocities, although due to its smaller size their results are not competitive with the results shown in Fig. \ref{fig:limit}.

\section{Summary and outlook \label{sub:outlook}}
We have described two searches using IceCube for cosmic magnetic monopoles 
for velocities $>\,0.51\,c$.
One analysis focused on high monopole velocities at the detector $v>0.76\,c$ where the mono\-pole produces Cherenkov light and the resulting detector signal is extremely bright. The other analysis considers lower velocities $>\,0.51\,c$ where the mono\-pole induces the emission of Cherenkov light in an indirect way and the brightness of the final signal is decreasing largely with lower velocity. Both analyses use geometrical information in addition to the velocity and brightness of signals to suppress background. The remaining events after all cuts were identified as background.
Finally the analyses bound the monopole flux to nearly two orders of magnitude below previous limits. 
Further details of these analyses are given in Refs. \cite{Posselt13, Obertacke15}. 

Comparable sensitivities are expected from the future KM3NeT instrumentation based on scaling the latest ANT\-ARES limit to a larger effective volume \cite{Km3net}. Also an ongoing ANT\-ARES analysis plans to use six years of data and estimates competitive sensitivities for highly relativistic velocities \cite{Antares15}. 

Even better sensitivities are expected from further years of data taking with IceCube, or from proposed volume extensions of the detector \cite{Gen2}. 
A promising way to extend the search to slower monopoles with $v \leq 0.5\,c$ is to investigate the luminescence they would generate in ice which may be detectable using the proposed low energy infill array PINGU \cite{PINGU14}.

\section*{Acknowledgments}

We acknowledge the support from the following agencies:
U.S. National Science Foundation-Office of Polar Programs,
U.S. National Science Foundation-Physics Division,
University of Wisconsin Alumni Research Foundation,
the Grid Laboratory Of Wisconsin (GLOW) grid infrastructure at the University of Wisconsin - Madison, the Open Science Grid (OSG) grid infrastructure;
U.S. Department of Energy, and National Energy Research Scientific Computing Center,
the Louisiana Optical Network Initiative (LONI) grid computing resources;
Natural Sciences and Engineering Research Council of Canada,
WestGrid and Compute/Calcul Canada;
Swedish Research Council,
Swedish Polar Research Secretariat,
Swedish National Infrastructure for Computing (SNIC),
and Knut and Alice Wallenberg Foundation, Sweden;
German Ministry for Education and Research (BMBF),
Deutsche Forschungsgemeinschaft (DFG),
Helmholtz Alliance for Astroparticle Physics (HAP),
Research Department of Plasmas with Complex Interactions (Bochum), Germany;
Fund for Scientific Research (FNRS-FWO),
FWO Odysseus programme,
Flanders Institute to encourage scientific and technological research in industry (IWT),
Belgian Federal Science Policy Office (Belspo);
University of Oxford, United Kingdom;
Marsden Fund, New Zealand;
Australian Research Council;
Japan Society for Promotion of Science (JSPS);
the Swiss National Science Foundation (SNSF), Switzerland;
National Research Foundation of Korea (NRF);
Danish National Research Foundation, Denmark (DNRF)

\section*{Appendix}

Table \ref{tab:limits} gives the numeric values of the derived limits of both analyses.
Tables \ref{tab:cutsIC40}, \ref{tab:cutsIC86} and \ref{tab:bdtIC86} show the event selection of both analyses in detail which illustrates how magnetic mono\-poles can be separated from background signals in IceCube.


\begin{table}[!h]
	
	\caption{Values of final limits of both analyses
	}
	\label{tab:limits}       
	\begin{tabular}{p{0.5cm}|p{0.8cm}p{2.0cm}|p{0.8cm}p{2.0cm}}
		\hline\noalign{\smallskip}
		Conf. & Velocity $v/c$ & $ \Phi_{90\%} \, / \, 10^{-18} $  [$cm^{-2}$ $s^{-1}$ $sr^{-1}$]  &
		Velocity $v/c$ & $ \Phi_{90\%} \, / \, 10^{-18} $  [$cm^{-2}$ $s^{-1}$ $sr^{-1}$]\\
		\noalign{\smallskip}\hline\noalign{\smallskip}
		
		IC40 &  0.76   & $7.73$ 
		& 0.8	   & $\,\,\,\,$$3.89$\\
		& 0.9	   & $3.06$
		& 0.995  
		& $\,\,\,\,$$2.90$ \\
		
		\noalign{\smallskip}\hline\noalign{\smallskip}
		
		IC86 & 0.510 &  8.71 
		& 0.517 &  $\,\,\,\,$7.58 \\
		& 0.523 &  6.71 
		& 0.530 &  $\,\,\,\,$6.02 \\
		& 0.537 &  5.49 
		& 0.543 &  $\,\,\,\,$4.33 \\
		& 0.550 &  3.54 
		& 0.557 &  $\,\,\,\,$3.01 \\
		& 0.563 &  2.66 
		& 0.570 &  $\,\,\,\,$2.38 \\
		& 0.577 &  2.18 
		& 0.583 &  $\,\,\,\,$2.05 \\
		& 0.590 &  1.94 
		& 0.597 &  $\,\,\,\,$1.86 \\
		& 0.603 &  1.80 
		& 0.610 &  $\,\,\,\,$1.75 \\
		& 0.617 &  1.70 
		& 0.623 &  $\,\,\,\,$1.65 \\
		& 0.630 &  1.62 
		& 0.637 &  $\,\,\,\,$1.59 \\
		& 0.643 &  1.57 
		& 0.650 &  $\,\,\,\,$1.56 \\
		& 0.657 &  1.56 
		& 0.663 &  $\,\,\,\,$1.55 \\
		& 0.670 &  1.55 
		& 0.677 &  $\,\,\,\,$1.55 \\
		& 0.683 &  1.54 
		& 0.690 &  $\,\,\,\,$1.56 \\
		& 0.697 &  1.57 
		& 0.703 &  $\,\,\,\,$1.58 \\
		& 0.710 &  1.59 
		& 0.717 &  $\,\,\,\,$1.59 \\
		& 0.723 &  1.59 
		& 0.730 &  $\,\,\,\,$1.58 \\
		& 0.737 &  1.58 
		& 0.743 &  $\,\,\,\,$1.59 \\
		& 0.750 &  1.94 
		& 0.757 &  $\,\,\,\,$2.29 \\
		& 0.763 &  2.65 
		& 0.770 &  $\,\,\,\,$3.02 \\
		& 0.777 &  3.39 
		& 0.783 &  $\,\,\,\,$3.10 \\
		& 0.790 &  2.81 
		& 0.797 &  $\,\,\,\,$2.54 \\
		& 0.803 &  2.67 
		& 0.810 &  $\,\,\,\,$3.23 \\
		& 0.817 &  4.14 
		& 0.823 &  $\,\,\,\,$5.28 \\
		& 0.830 &  6.84 
		& 0.837 &  $\,\,\,\,$7.85 \\
		& 0.843 &  7.97 
		& 0.850 &  $\,\,\,\,$8.77 \\
		& 0.857 &  9.05 
		& 0.863 &  $\,\,\,\,$8.82 \\
		& 0.870 &  8.61 
		& 0.877 &  10.39 \\
		
		\noalign{\smallskip}\hline
	\end{tabular}
\end{table}

\clearpage

\begin{table*}
	
	\caption{Description of all  cuts  in the highly relativistic analysis. 
		For some cuts only the 10\% of the DOMs with the highest charge (HC) were chosen
	}
	\label{tab:cutsIC40}       
	\begin{tabular}{p{1.5cm}p{3.0cm}p{0.5cm}p{5cm}p{5.3cm}}
		\hline\noalign{\smallskip}
		Cut \newline Variable & Cut value & Hits & Description & Motivation \\
		\noalign{\smallskip}\hline\noalign{\smallskip}
		
		$n_{\textrm{DOM}}$              & $>60$                   & all & Number of hit DOMs & Improve quality of $n_{\textrm{NPE}}/n_{\textrm{DOM}}$ variable\\
		$n_{\textrm{NPE}}/n_{\textrm{DOM}}$ & $\geq 8$                & all & Average number of photo-electrons per DOM & Reduce events with low relative brightness\\
		$v$                             & $\geq 0.72\, c$            & HC  & Reconstructed velocity & Reduce cascade events\\
		$n_{\textrm{String}}$           & $\geq 2$                & HC  & Number of hit strings & Reduce cascade events\\
		$t$                             & $\geq 792\;\textrm{ns}$ & HC  & Time length of an event; calculated by ordering all hits in time and subtracting the last minus the first time value & Reduce cascade events\\
		Topological \newline Trigger    & no split                & all & Attempt to sort the hits in an event into topologically connected sets & Split coincident events\\
		$\textrm{NHF}_{100}$            & $< 0.784$               & all & Fraction of DOMs with no hit in a $100\;\textrm{m}$ cylinder radius around the reconstructed track & Reduce (coincident/noise) events with spurious reconstruction\\
		$d_{\bot}$                      & $<(110 - 64\cdot\textrm{NHF}_{100})\;\textrm{m}$ & all & Root mean square of the lateral distance of hit DOMs (weighted with DOM charge) from the track & Reduce (coincident/noise) events with spurious reconstruction\\
		$d_{\textrm{Gap 100}}$          & $\leq 420\;\textrm{m}$  & all & The maximal length of the track, which got no hits within the specified track cylinder radius in meters & Reduce (coincident/noise) events with spurious reconstruction\\
		
		\noalign{\smallskip}\hline\noalign{\smallskip}
		\multicolumn{5}{c}{Low Brightness Cuts ($n_{\textrm{NPE}}/n_{\textrm{DOM}}< 31.62$)}\\
		\noalign{\smallskip}\hline\noalign{\smallskip}
		
		$t$                             & $> 1400\;\textrm{ns}$   & HC  & See above & See above (hardened cut)\\
		$d_{\textrm{Gap 100}}$          & $> 112\;\textrm{m}$ and $< 261\;\textrm{m}$ & all & See above & See above (hardened cut)\\
		$\cos\theta$                    & $< -0.2$                & HC  & Reconstructed zenith angle & Reduce events caused by mostly downward moving air shower muons\\
		
		\noalign{\smallskip}\hline\noalign{\smallskip}
		\multicolumn{5}{c}{High Brightness Cuts ($n_{\textrm{NPE}}/n_{\textrm{DOM}}\geq 31.62$)}\\
		\noalign{\smallskip}\hline\noalign{\smallskip}
		
		$t$                             & $\geq (792 + 2500\cdot\cos\theta)\;\textrm{ns}$ & HC & See above & Reduce events caused by mostly downward moving air shower muons (supportive cut)\\
		$n_{\textrm{NPE}}/n_{\textrm{DOM}}$ & $\geq 31.62 + 330\cdot\cos\theta$ & all & See above & Reduce events caused by mostly downward moving air shower muons\\
		
		\noalign{\smallskip}\hline
	\end{tabular}
\end{table*}

\begin{table*}
	
	\caption{Description of all cuts in the mildly relativistic analysis and the according event rate}
	\label{tab:cutsIC86}       
	\begin{tabular}{p{1cm}p{1.3cm}p{1.3cm}p{5.8cm}p{5.8cm}}
		\hline\noalign{\smallskip}
		
		Cut \newline Variable  & Cut value & Data Rate [Hz] & Description & Motivation \\
		\noalign{\smallskip}\hline\noalign{\smallskip}
		
		$\theta$ & $\geq 86^{\circ}$ & $ 2.30 \cdot 10^{1}$ & Reconstructed zenith angle using improved Line\-Fit 
		& Reduce muons from air showers which are  significantly reduced at this angle because of the thick atmosphere and ice; this also requires a cut on the successful fit-status of the reconstruction \\
		$v$ & $\leq 0.83\, \, c$ &   & Reconstructed velocity & Only used in training to focus on low velocities\\
		$n_{\textrm{String}}$ & $\geq 2$ &  $ 1.86 \cdot 10^{1}$ & Number of hit strings & Improve data quality and reduce pure noise events \\
		$n_{\textrm{DOM}}$ & $\geq 6$ &  $ 1.64 \cdot 10^{1}$ & Number of hit DOMs & Improve data quality and reduce pure noise events \\
		$d_{\textrm{Gap 100}}$ & $\leq 300\;\textrm{m}$ &  $ 1.41 \cdot 10^{1}$ & The maximal track length of the track, which got no hits within the specified track cylinder radius in meters & Reduce coincident events and noise events \\
		$d_{\textrm{Separation}}$ & $\geq 350\;\textrm{m}$ &  $ 2.62\cdot 10^{-1}$ & The distance the Center-of-Gravity (CoG) positions of the first and the last quartile of the hits, within the specified track cylinder radius, are separated from each other. & Reduce down-going events, corner-clippers, and cascades \\
		$z_{\textrm{CoG}}$ & $\geq -400\; \textrm{m}$ &  $ 2.40\cdot 10^{-1}$ & The z value of the position of the  CoG of the event. & Reduce  horizontally mis-reconstructed high energy tracks at the bottom of the detector \\
		$z_{\textrm{DOM}}$ & & & height $z$ of the position of a certain DOM & \\
		$z_{\textrm{travel}}$ & $\geq 0 \, \textrm{m}$ &  $ 1.30\cdot 10^{-1}$ & Average penetration depth of hits defined from below: The average over ($z_{\textrm{DOM}}$ minus the average over the $z_{\textrm{DOM}}$ values of the first quartile of all hits) & Reduce coincident events, down-going tracks and cascades \\
		BDT Score & $ \geq 0.47$  &  $ 1.12 \cdot 10^{-7}$ & Score reaching from $-1$ to 1 representing how signal-like an event is & For the choice of the value see text; see Tab. \ref{tab:bdtIC86} for the used variables \\
		
			\noalign{\smallskip}\hline
		\end{tabular}
	
\end{table*}

\begin{table*}
	
	\caption{Description of the variables used in the BDTs of the mildly relativistic analysis}
	\label{tab:bdtIC86}       
	
	\begin{tabular}{p{1.2cm}p{1cm}p{13.9cm}}
		
		\hline\noalign{\smallskip}
		
		 mRMR \newline Importance & BDT \newline Variable & Description  \\
		
		\noalign{\smallskip}\hline\noalign{\smallskip}
		
		1 & $n_{\textrm{DOM 100}}$ & The number of hit DOMs within the specified cylinder radius in meters around the reconstructed track \\
		2 & $\bar{s}$  & The mean of all distances of hits from the reconstructed track \\
		3 & $t_{\textrm{Gap}}$  & Largest time gap between all hits ordered by time\\
		4 & $d_{\textrm{Gap 100}}$   & See above \\
		5 & $d_{\textrm{Separation}}$  & See above\\
		6 & $\bar{s}_{\textrm{NPE}}$  & The average DOM distance from the track weighted by the total charge of each DOM \\
		7 & $n^*_{\textrm{DOM 50}}$  &  The number of  DOMs with no hit within the specified cylinder radius in meters around the reconstructed track \\
		8 & $z_{\textrm{travel}}$ & See above   \\
		9 & $z_{\textrm{pattern}}$  & All hits are ordered in time. If a DOM position of a pulse is higher than the previous $z_{\textrm{pattern}}$  increases with +1. If the second pulse is located lower in the detector $z_{\textrm{pattern}}$  decreases with -1. So this variable gives a tendency of the direction of a track  \\
		10 & $n_{\textrm{DOM 50}}$  & The number of hit DOMs within the specified cylinder radius in meters around the reconstructed track  \\
		11 & $v$ &  See above  \\
		12 & $k_{\textrm{100}}$ &  The smoothness values reaching from $-$1 to 1 how smooth the hits are distributed within the specified cylinder radius around the reconstructed track \\
		13 & $t_w$ &  The weighted deviation of all hit times from the charge weighted mean of all hit times distribution \\
		14 & $t$ &  Time length of an event; calculated by ordering all hits in time and subtracting the last minus the first time value   \\
		15 & $\bar{z}_{\textrm{DOM}}$ &  Mean of all $z_{\textrm{DOM}}$ per event\\

		\noalign{\smallskip}\hline
	\end{tabular}
\end{table*}

\clearpage


\bibliographystyle{spphys}       
\bibliography{Lib.bib}


\end{document}